\documentstyle[preprint,aps]{revtex}
\newcommand{\nc}{\newcommand}
\nc{\beqn}[1]{\begin{equation} \label{Eqn:#1}}
\nc{\eeqn}{\end{equation}}
\nc{\newsect}[2]{\section{#2}\label{Sec:#1}}
\nc{\eq}[1]{(\ref{Eqn:#1})}
\nc{\cit}[1]{\cite{#1}}
\nc{\sect}[1]{\ref{Sec:#1}}
\nc{\bib}[1]{\bibitem{#1}}
\nc{\e}[1]{\/{\em #1\/}}
\nc{\ie}{\e{i.e.}}
\nc{\eg}{\e{e.g.}}
\nc{\etal}{\e{et al.}}
\nc{\apriori}{\e{a priori}}
\nc{\coord}{co\"{o}rdinate}
\nc{\cross}{\times}
\nc{\vdot}{\cdot\!}
\nc{\Fdual}{\widetilde{F}}
\nc{\littlefrac}[2]{\case{#1}/{#2}}
\nc{\half}{\littlefrac{1}{2}}
\nc{\paper}[5]{#1, {\it #2\/}\ {\bf #3} (#4) #5}
\nc{\book}[7]{#1, {\it #2\/}#3\ (#4, #5, #6)#7}
\nc{\booknocity}[6]{#1, {\it #2\/}#3\ (#4, #5)#6}
\nc{\PRx}[5]{\paper{#2}{Phys. Rev. #1}{#3}{#4}{#5}}
\nc{\PLx}[5]{\paper{#2}{Phys. Lett. #1}{#3}{#4}{#5}}
\nc{\al}{\alpha}
\nc{\be}{\beta}
\nc{\de}{\delta}
\nc{\g}{\gamma}
\nc{\G}{\Gamma}
\nc{\k}{\kappa}
\nc{\la}{\lambda}
\nc{\La}{\Lambda}
\nc{\m}{\mu}
\nc{\n}{\nu}
\nc{\s}{\sigma}
\nc{\z}{\zeta}
\nc{\eps}{\varepsilon}
\nc{\pard}{\partial}
\nc{\id}{\equiv}
\nc{\nline}{\nonumber \\}
\nc{\paren}[1]{\left( #1 \right)}
\nc{\brac}[1]{\left[ #1 \right]}
\nc{\braces}[1]{\left\{ #1 \right\}}
\nc{\vect}[1]{\mbox{\boldmath{$#1$}}}
\nc{\gcap}[1]{{\it #1}}
\nc{\modsign}[1]{\left| #1 \right|}
\nc{\del}{\gcap{\vect{\nabla}}}
\nc{\dummy}{\mbox{}}
\nc{\spacer}{\mbox{ }\mbox{ }}
\nc{\vmu}{\vect{\m}}
\nc{\vmudot}{\dot{\vect{\m}}}
\nc{\vB}{\vect{B}}
\nc{\vE}{\vect{E}}
\nc{\vv}{\vect{v}}
\nc{\vs}{\vect{s}}
\nc{\BMT}{Bargmann--Michel--Telegdi}
\nc{\TBMT}{Thomas--\BMT}
\nc{\EL}{Euler--Lagrange}
\nc{\Hrasko}{Hrask\'{o}}
\nc{\geff}{g_{\text{eff}}}
\nc{\vOm}{\vect{\Omega}}
\begin{document}
\preprint{UM--P--92/93}
\title{Electromagnetic Deflection of Spinning Particles}
\author{John P. Costella and Bruce H. J. McKellar}
\address{
School of Physics, The University of Melbourne,
Parkville, Victoria 3052, Australia
}
\date{September 1992}
\maketitle
\begin{abstract}
We show that it is possible to
obtain self-consistent and physically acceptable
relativistic classical equations of motion for a point-like spin-half particle
possessing an electric charge and a magnetic dipole moment,
directly from a manifestly covariant Lagrangian, if the classical degrees
of freedom are appropriately chosen.
It is shown that the equations obtained encompass the well-tested
Lorentz force and \TBMT\ spin
equations, as well as providing a definite specification of the classical
\e{magnetic dipole force}, whose exact form has been the
subject of recent debate.
Radiation reaction---the force and torque on an accelerated particle
due to its self-interaction---is neglected at this stage.
\end{abstract}
\narrowtext

\newsect{Intro}{Introduction}
The ``classical limit'' of mechanics has always played an important
role in practical physics.
While it may be regarded as preferable to describe the behaviour
of a physical system by solving the problem exactly within quantum mechanics,
in practice one need not always do so.
If one is only interested in \e{expectation values} of operators,
then one can make good use of Ehrenfest's theorem---which holds good
under quite general circumstances (see, \eg, \cit{Dirac58}, \S31)---and
instead solve the equations of \e{classical} Hamiltonian or Lagrangian
dynamics.
Thus, for example,
the Lorentz force law for charged particles, and the
\TBMT\ spin precession equation \cit{Thomas27,Bargmann59} for
particles with spin, are used to advantage every day---implicitly
or explicitly---in a wide variety of practical situations.

These two examples, however, owe their widespread acceptance
largely to an important quality they possess: they can be derived
simply from a knowledge of the external electromagnetic multipolar
fields of their point-like sources, \e{regardless} of the detailed structure
of these sources \cit{Hrasko71}.
On the other hand,
the \e{force} on a point-like particle with spin is \e{not} uniquely defined
by the external properties of its dipolar electromagnetic fields;
this fact, well noted by Thomas in 1927 \cit{Thomas27}, and more
recently by \Hrasko\ \cit{Hrasko71}, is not yet
universally appreciated.
To obtain the ``correct'' force on (and, hence, complete equations
of motion for) a particle with a magnetic moment,
one must therefore make \e{some} assumptions,
explicit or otherwise, about the nature of the object creating the dipolar
field;
it is this freedom that has contributed to the ongoing controversy
on the subject \cit{Barone73,Boyer87,Aharonov88,Goldhaber89,%
Aharonov84}.

It would be impossible to review here all of the
various assumptions underlying
previous attempts at a complete classical description of particles
with spin;
instead, we shall simply state the premises and results
of a theoretical analysis that we have
carried out.
In section~\sect{Lagrangian}, we outline our considerations in choosing
a suitable relativistic classical Lagrangian for a particle with spin.
We then, in section~\sect{Derivation}, outline our derivation of the equations
of motion from this Lagrangian, and present the results in what we believe
are the most transparent forms.
As will be seen, the technique we shall use leads to the
omission of all aspects of radiation reaction---which is an omission
that should be repaired---but nevertheless
the results obtained do allow extensive contact with existing knowledge
about the classical behaviour of particles with spin.

\newsect{Lagrangian}{An Appropriate Lagrangian}
As already mentioned, some non-trivial assumptions must be made
for one to obtain the force equation of motion for even a point-like
particle endowed with a magnetic dipole moment.
Our approach is to use a Lagrangian approach from the outset, and
hence benefit from the abovementioned guarantee that Ehrenfest's
theorem provides.
This route is, however, still a potential minefield; as recently
emphasised by
Barut and Unal \cit{Barut89}, the ``spinning top''
(or ``current loop'') phase space degrees of freedom
are \e{not} appropriate to quantum
spin-half particles.
We shall return to this question shortly.

Our first consideration is to outline precisely \e{what type} of
system we wish to address.
In this specification we shall be exacting: we shall only be
considering the \e{intrinsic} magnetic dipole moment of a
spin-half, pointlike, structureless particle---together
with any electric charge interaction it possesses, of course---in
externally applied (but, in general, time- and space-varying)
electromagnetic fields.
It is only with such a tight restriction of our focus that we
can proceed confidently in the classical realm at all, as will become
apparent.
For definiteness, we consider the \e{electron}
a good approximation to the type of particle we are considering;
however, we shall not claim any predictive control over
contributions from the
\e{anomalous} magnetic moment of the real electron,
arising as it does from QED and other processes that are, in reality,
additional
degrees of freedom over those properly described
by the single-particle Dirac equation.
We shall not enter the debate as to whether a particle
might ``intrinsically'' have a g-factor differing from 2 in the
single-particle Dirac equation; if so, then such
``anomalous'' moments would be included in our considerations.
In any case, since the pure Dirac moment of the electron dominates
its anomalous moment numerically, our considerations will, at least,
provide useful practical results for real electrons to leading order
in the fine-structure constant.

Our main task in this section is to explain the
particular Lagrangian that we have chosen
to represent the magnetic interaction of such a particle.
There are two distinct parts to this decision: firstly, the choice of the
\e{value} that the Lagrangian should take; and, secondly, a
determination of the appropriate classical \e{degrees of freedom}
that are to be used in the \EL\ equations---and, hence, the functional
form of the Lagrangian.
The former choice is laid out for us: the Dirac
equation gives us
\beqn{FirstLag}
L_{{\text{int}}} = -\frac{g q \hbar}{8m} \sigma^{\al\be}\!F_{\al\be},
\eeqn
where $q$ is the charge of the particle ($=-e$ for an electron) and
$g$ is its gyromagnetic ratio ($=2$ for the simple minimally-coupled
Dirac equation), and our units follow SI conventions with the
exception that we set $\eps_0=\m_0=c=1$.
However, to entertain the possibility of the single-particle Dirac
equation fully
describing a \e{neutral} particle with an anomalous moment, we shall
replace the $g$ factor by the corresponding magnetic moment
$\mu = gq\hbar/4m$, thereby separating the interaction of electric
charge from that of magnetic moment, at least formally.

Our second task is infinitely more perilous: choosing an appropriate
set of generalised \coord s to represent a spin-half particle.
Firstly, we observe that the \e{position} of a particle is
usually assumed to be an appropriate quantity in the classical
limit; we shall also make this assumption.
Thus, we take the (expectation value of the) four-position $z^\al(\tau)$ to be
four of the independent degrees of freedom.
(One can show that, in a general dynamical framework allowing particles
with time-dependent masses, all four of these \coord s are
indeed independent.)
If we take the proper time for the particle, $\tau$, to be the
generalised time of the classical Lagrangian framework, then
the generalised velocities corresponding to the $z^\al$
are given by
\beqn{UDefn}
U^\al \id \frac{dz^\al}{d\tau} \id \dot{z}^\al .
\eeqn
Having defined the four-velocity thus,
we can then simply define the
\e{centre of energy frame} for the particle as that
instantaneous Lorentz frame in which $U^\al=(1,0,0,0)$.

We now turn to the \e{magnetic moment} of the particle.
For a spin-half particle, the magnetic moment is parallel to the
spin.
In the centre of energy frame, we expect the particle's spin to
be describable by a two-component spinor, of fixed magnitude
$s=\half \hbar$.
The expectation value of this spinor is equivalent to a
\e{fixed-length three-vector}, in this frame, whose direction
describes simultaneously the expectation value of polarisation
along any arbitrary axis, as well as the phase difference between the
``up'' and ``down'' components along such an axis, but which
throws away the \e{overall} phase of the wavefunction (as does
any reduction to the classical limit).
It is this \e{fixed-length} three-vector that we shall assume to
represent the magnetic moment of the particle.

Consider now
the \e{dynamical} nature of
spin angular momentum in the classical limit. It would seem, from
the previous paragraph, that the spin
should \e{also} be considered as a fixed-length three-vector in the
centre of energy frame of the particle.
However, this would preclude writing down the usual classical
kinetic Lagrangian of rotational motion in terms of
generalised \e{velocities} (\ie\ the time derivatives of the Euler
angles), namely, in the non-relativistic limit,
 \cit{Goldstein80}
\beqn{SpinKineticLag}
L=\half\, \vect{\omega} \cdot \vect{I} \cdot \vect{\omega},
\eeqn
and thereby
preclude obtaining the \e{torque} on the particle by means
of the \EL\ equations for the rotational degrees of freedom.
Our approach, therefore, shall be the following: we shall
consider the \e{spin} of the particle to be given, in the
non-relativistic limit, by its usual form
\[ \vect{s} = \vect{I} \cdot \vect{\omega}, \]
with the ``kinetic'' term \eq{SpinKineticLag} retained in
the Lagrangian in this same limit.
This allows the \e{magnitude} of the spin,
$\modsign{\vect{s}}=s$ to be an \apriori\ dynamical variable.
(We know, of course, that $s$ should, ultimately, be a
constant of the motion (\ie\ $\half\hbar$)
if we are to accept the equations of
motion as applicable to spin-half particles; we shall discuss this
necessity in greater detail shortly.)
We shall, on the other hand,
still consider the \e{magnetic moment} of the particle
to be formally a function of
the generalised \coord s themselves (\ie\ the Euler angles)
and \e{not} the velocities.
This distinction between the spin and magnetic moment of an
electron is an unfamiliar concept within quantum mechanics;
indeed, in that situation, no distinction need be made.
However, this distinction is vital in the \e{classical} framework---at
least, if
one wants to derive the equations of motion from a Lagrangian---as
the spin angular momentum and magnetic dipole moment of
an arbitrary classical object are not, in general, related in any
special way.
Our ansatz that the magnetic moment is dependent on the Euler angles
alone cannot be justified \apriori;
the only justification will be that it produces equations of motion that
satisify the strict requirements of consistency for spin--half particles---that
have not, to our knowledge, been satisfied by previous classical
Lagrangian approaches.
We further note, at this stage, that the ultimate parallelism of spin and
magnetic moment \e{does not} bar us from considering them to have
different functional dependencies: the fact that they are parallel may be
introduced, in this classical context, as simply a ``constitutive'' relation,
which can only be fully justified upon investigation of the quantum
mechanical analysis.

We now turn to synthesising the information that we have laid out above
in a self-consistent, relativistic, Lagrangian framework. Firstly, it is
necessary to generalise the non-relativistic concept of the spin
angular momentum $\vect{s}$ to the relativistic domain.
This procedure is carried out in many textbooks on electrodynamics
(see, \eg, \cite{Jackson75}, pp.\ 556--560); we shall not add anything
new.
One simply defines a four-vector $S^\al$ such that, \e{in the rest
frame} of the particle, it has vanishing zero-component $S^0$, and
its three-vector part
$\vect{S}$ is equal to the non-relativistic spin $\vect{s}$.
Since, by definition, the three-vector part of the four-velocity, $\vect{U}$,
vanishes in this frame, the identity
\beqn{SpinUOrthog}
S^\al U_\al = 0
\eeqn
shows that only three of the $S^\al$ are independent, as one would
expect from the non-relativistic case.

In a completely analogous way, one can generalise the non-relativistic
angular velocity vector $\vect{\omega}$ and the moment of inertia
tensor $\vect{I}$ to their relativistic counterparts $\omega^\al$ and
$I^{\al\be}$.
The non-relativistic rotational ``kinetic'' Lagrangian \eq{SpinKineticLag}
can then be written relativistically as
\beqn{SKinLagRel}
L_{\text{rot}}=\half \omega^\al I_{\al\be}\, \omega^\be.
\eeqn
One might wonder, at this point, how a spin-half particle can have
a ``kinetic'' term of rotation, when it is well known that there is
\e{no} classical ``rotating model'' that represents spin angular
momentum.
The answer is as follows: the \e{functional form} of the spin
kinetic Lagrangian must be retained in the classical limit,
regardless of whether or not it corresponds to any particular
``model'' that one might dream up.
In fact, it will be found that the quantities $\omega^\al$ and
$I_{\al\be}$ will
 \e{completely disappear} from the final
equations of motion; the only remaining quantity will be the
\e{physically observable} quantity $S^\al$.
Thus, by what might arguably be considered
a sleight of hand, we can analyse the system in question
within Lagrangian mechanics, without recourse to
the particular Newtonian models of refs
\cit{Hrasko71,Boyer87,Aharonov88,Goldhaber89}, nor the
original heuristic (albeit brilliant) arguments of Thomas
\cit{Thomas27} and Bargmann, Michel and Telegdi
\cit{Bargmann59}.

Before we can write down the final expression for the
Lagrangian we shall use,
we must first ``massage'' the magnetic interaction
Lagrangian \eq{FirstLag} into a more suitable form.
Our major task is to the interpret the spin \e{tensor}, $\sigma^{\al\be}$,
that the Dirac equation introduces.
This is not a trivial task: translating between the spin \e{vector}
that we have already defined, $S^\al$, and a spin \e{tensor},
requires the use of both the alternating tensor $\eps^{\al\be\m\n}$,
and another four-vector.
Often, in quantum mechanics, one uses the \e{canonical momentum}
vector, $p^\al$, for this purpose; the resultant ``spin'' vector is known as
the \e{Pauli--Lubanski vector},
\beqn{PauliLub}
W^\al \id \eps^{\al\be\m\n}p_\be S_{\m\n}.
\eeqn
This quantity is, indeed, very useful in many situations.
However, we are here considering \e{Lagrangian} mechanics;
therefore, we should expect that \e{mechanical} momenta
(or, in other words, generalised velocities) should be
employed; \e{canonical} momenta belong to Hamiltonian
dynamics.
We therefore follow Jackson (\cit{Jackson75}, p.\ 556) in
using the \e{four-velocity} to define the transformation between the
spin tensor and the spin vector, namely
\beqn{SpinVectorId}
S^\al = \half \eps^{\al\be\m\n} U_\be S_{\m\n}.
\eeqn
It is straightforward to verify that the reverse transformation of
\eq{SpinVectorId}
is given by
\beqn{SpinTensorId}
S^{\al\be}=\eps^{\al\be\m\n}U_\m S_\n.
\eeqn
Insering this into \eq{FirstLag} and simplifying,
we finally obtain our desired magnetic interaction Lagrangian,
\beqn{FinalMagIntLag}
L_{\text{int}}=\m^\al \Fdual_{\al\be} U^\be,
\eeqn
where $\Fdual_{\al\be}\id\half{\eps_{\al\be}}^{\m\n}F_{\m\n}$ is
the dual electromagnetic field strength tensor, and
the magnetic moment four-vector, $\m^\al$ is,
as noted previously, considered to be
a four-vector ``embedded'' in the instrinsic rotational \coord s of the
particle, and hence is functionally dependent on the Euler angles,
but \e{not} their derivatives.

Our complete Lagrangian is then assembled simply from the
kinetic rotational term \eq{SKinLagRel}, the magnetic interaction term
\eq{FinalMagIntLag}, and the standard translational kinetic and
electric charge interaction terms:
\beqn{FinalLag}
L=\half m U^\al U_\al +  \half \omega^\al I_{\al\be}\, \omega^\be
+ q U^\al\! A_\al + \m^\al \Fdual_{\al\be} U^\be.
\eeqn
It is this Lagrangian, and most particularly its functional form,
that forms the basis of the following section.

\newsect{Derivation}{Derivation}
We now turn to the question of  deriving, from the \EL\ equations, the
equations of motion for the particle under study.
(In the following, our language shall describe classical
quantities in the same way that Newton would have done, but in
reality we are referring to the \e{expectation values}
of the corresponding quantum mechanical operators for the
single particle in question.)
The generalised \coord s for the particle, following the discussion
of the previous section, are taken to be
the four translational degrees of
freedom $z^\al$, together with the three
Euler angles describing the intrinsic ``orientation''
of the particle.

The mathematical manipulations necessary to obtain the
seven \EL\ equations are in principle
no different to those in everyday classical
mechanical problems \cit{Goldstein80}.
There is one subtlety, however, that enters into one's
consideration of the spin vector $S^\al$ and the magnetic moment
vector $\mu^\al$.
By their very definition, these vectors are ``tied'' to the particle's
four-velocity,
in the sense that identities such as \eq{SpinUOrthog} always holds
true.
One must therefore be careful when defining their proper-time
derivative: since they are, in effect, defined with respect to an
\e{accelerated} frame of reference, there is a difference
between taking the time-derivatives of the components of the
vector, and the components of the time-derivatives of the vector,
as General Relativity teaches us.
In fact, the philosophical framework of General Relativity tells
us that it is the \e{latter} that is the generally invariant,
``covariant'' derivative which should be used in the relativistic \EL\ %
equations; the former is simply the ``partial'' derivative.
However, it is straightforward to verify that they can be related via
\beqn{ProperTDeriv}
\frac{d}{d\tau} \paren{C^\al} \id \dot{C}^\al
+ U^\al \paren{\dot{U}_\be C^\be},
\eeqn
where $C^\al$ is \e{any}
space-like vector that is orthogonal to the particle's four-velocity
(such as $S^\al$ and $\m^\al$),
the left hand side denotes the ``covariant'' derivative,
and $\dot{C}^\al$ the ``partial'' derivative, of such a vector.
(The \e{Thomas precession} \cit{Thomas26,Thomas27} is, in fact,
just another way of expressing this ``General Relativity''
effect---namely, the non-vanishing commutator of Lorentz boosts.)

It is now relatively straightforward to
apply the \EL\ equations to the Lagrangian \eq{FinalLag}.
The three Euler angle degrees of freedom lead, just as in the
non-relativistic case \cit{Goldstein80}, to the \e{torque}
equation of motion for the particle.
With some algebra and simplification,
they (together with the identity \eq{SpinUOrthog} and the
relation \eq{ProperTDeriv})
lead to the four-vector equation of motion
\beqn{BMT}
\dot{S}_\al + U_\al \dot{U}^\be S_\be = F_{\al\be} \m^\be
  +  U_\al \m^\n F_{\n\be} U^\be.
\eeqn
It should come as no surprise that this equation is precisely
that obtained by Bargmann, Michel and Telegdi (equation~(6) of
ref.\ \cit{Bargmann59}), \e{before} they simply substituted in the
Lorentz force law for $\dot{U}_\be$ (\ie\ equation~(7) of
\cit{Bargmann59}).
As was noted in section~\sect{Intro}, this
equation---\eq{BMT}---is true
in general for \e{any} point-like object generating an external
magnetic dipole field \cit{Hrasko71}.

We now turn our attention to the
four \e{translational} degrees of freedom of the particle.
The \EL\ equations for these \coord s, for the \e{same}
Lagrangian \eq{FinalLag} as used above to generate the
\BMT\ equation, immediately yield the
four-equation of motion
\beqn{RelFirstForce}
\frac{d}{d\tau} \paren{mU_\al} = qF_{\al\be} U^\be
  + U^\be \m^\n \pard_\n \Fdual_{\al\be}
  + \Fdual_{\al\be} \paren{\dot{\m}^\be + U^\be \dot{U}_\n \m^\n }
  + \eps_{\al\be\m\n}\m^\be U^\n J^\m_{\text{ext}}.
\eeqn
The first term on the right side is, of course, simply the Lorentz force.
The second and third terms, on the other hand,
are infinitely more interesting. The second term
represents the \e{gradient} forces on the magnetic dipole moment,
 \ie, in the non-relativistic limit, $\paren{\vmu \vdot \del} \vB$.
The third term represents a force of the type $-\dot{\vmu} \cross \vE$
in the non-relativistic limit.
The reason that they are so interesting is that
many authors have argued that \e{precisely these terms} should
constitute the non-relativistic limit of the magnetic dipole force
(see, \eg, \cit{Barone73} and references therein; also
\cit{Hrasko71,Aharonov88,Goldhaber89}).
Here, we have obtained these terms from a relativistic Lagrangian directly,
without need for assumptions other than those outlined in the previous
section.

The last term in \eq{RelFirstForce} is, as a matter of principle, much
deeper, but in practice completely ignorable.
It constitutes a \e{contact force} between the particle and the
electromagnetic current that is generating the ``external'' field.
In practice, such an interaction is usually ignored;
however, upon further investigation, it is recognised that
it is precisely this
contact force that allows the expression \eq{RelFirstForce} to
otherwise resemble so closely the force on a ``monopole-constructed''
dipole model \cit{Hrasko71,Boyer87}, while on the other hand
possessing an interaction energy equivalent to a ``current loop''
model \cit{Hrasko71,Boyer87,Aharonov88,Goldhaber89}
as required for agreement with atomic hyperfine levels \cit{Opat76}.
It is in this way that the ``Lagrangian-based'' model of this paper
seems to select those
desirable properties of \e{both} the ``monopole-constructed'' and
``current-loop'' models, while being classically equivalent to neither.

Now that we have dealt with the \e{dynamics} of the  \EL\ formalism,
and have obtained allegedly appropriate equations of motion,
we can now specify the exact ``constitutive relation'' between the
magnetic moment and the spin of the particle, for the particular case
we wish to study: a spin-half particle.
{}From quantum mechanics, we know that the these two quantities are
always parallel, namely,
\beqn{Parallel}
\m^\al = \zeta S^\al,
\eeqn
where $\zeta\id\m/s$ is some constant, a property of the particle in
question, which,
for particle of charge $q$, is commonly written in terms of the $g$ factor
as $\zeta\id gq/2m$.
Before we can use \eq{Parallel}, however, we first note that
equation \eq{RelFirstForce} is itself in a somewhat
awkward form: the right hand side involves both
$\dot{\m}^\al$ and $\dot{U}^\al$.
However, the parallelism condition \eq{Parallel}
allows us to \e{uncouple} equations \eq{BMT} and
\eq{RelFirstForce} simply by \e{substituting} \eq{BMT}
into \eq{RelFirstForce}, yielding
\beqn{Uncoupled}
\frac{d}{d\tau} \paren{mU_\al} = qF_{\al\be} U^\be
  + U^\be \m^\n \pard_\n \Fdual_{\al\be}
  + \zeta \braces{ \Fdual_{\al\be} {F^\be}_\n \m^\n
    + \Fdual_{\al\be} U^\be \paren{\m^\sigma F_{\sigma\tau} U^\tau} },
\eeqn
where we have, for practical simplicity, dropped the
contact term in \eq{RelFirstForce}.

There are now several important things we can do with
equations \eq{BMT} and \eq{Uncoupled}.
Most importantly, we examine how the \e{mass} of the particle
changes with time; if it is not constant, then our equations of motion
cannot possibly apply to any particle, such as an electron or muon,
that has a constant rest mass.
(It should be noted that the classical relativistic formalism we have
employed has made no assumptions as to the constancy of the
rest mass $m$ with time: in general, the mass may change as the system in
question gains energy from or loses energy to the external field.)
It is straightforward to verify that the general proper-time rate of
change of the mass of a particle, $\dot{m}$, can be expressed as
\beqn{MassDeriv}
\dot{m} \id U^\al \frac{d}{d\tau} \paren{m U_\al }.
\eeqn
Using the the identity \eq{SpinUOrthog}, and the
fact that $\Fdual_{\al\be} {F^\be}_\n$ is proportional
to the metric, $g_{\al\n}$,
it follows that \eq{Uncoupled} yields, in \eq{MassDeriv},
$\dot{m}=0$. In other words, the set of equations \eq{BMT} and
\eq{Uncoupled} \e{rigorously maintain} the constancy of
the mass of the particle.
This property is far from trivial; for example, in a classic
textbook (\cit{Barut64}, p.\ 74),
a time-varying ``effective mass'' of the
electron was introduced,
due to the fact that the equations of motion derived therein
allowed \e{time-changing} rest masses.
It is of utmost importance that this difficulty---present in most
previous attempts at consistent equations of motion for a dipole---is
overcome in the Lagrangian treatment of this paper.

We now verify that the the parallelism condition, \eq{Parallel}, is
consistent with the equations of motion \eq{BMT} and \eq{Uncoupled}.
The reason for this concern is that we have assumed that the magnetic
moment is a
constant-magnitude vector, whereas the spin itself is a dynamical quantity
that may, in general, \e{change} its magnitude.
We must verify that, if \eq{Parallel} is assumed to hold true at one
particular proper time $\tau$, the equation of motion \eq{BMT}
does not change the magnitude of the spin.
On diffentiation of the identity $s^2 \id -S^\al S_\al$,
one finds $\dot{s}=-\half \dot{S}^\al S_\al$; use of \eq{Parallel} in
\eq{BMT} then shows that \eq{BMT} does, in fact, yield
$\dot{s}=0$.
Thus, the magnitude of the spin is a rigorous constant of the
motion, just as is the mass.

Our work is now essentially complete.
However, to make \e{practical} use of the equations \eq{BMT} and
\eq{Uncoupled}, it is appropriate to both present them in a
more computationally-friendly form, and to highlight clearly where they
add to existing knowledge of spin-half particles.
These two tasks can essentially be carried out simultaneously.
We shall transform \eq{BMT} and \eq{Uncoupled} into the
standard ``$3+1$'' form, as, for example, presented in Jackson's textbook
(\cit{Jackson75}, pp.\ 556-560).
This involves using the three-velocity of the particle in some \e{particular}
fixed ``lab'' frame, $\vv$ (Jackson uses the symbol $\vect{\be}$), as well
as the three-spin
$\vs$ of the particle as seen in its rest frame, but referred to
the (non-rotating)
\coord s of the ``lab'' frame.
The procedure used to effect this transformation is described in detail in
\cite{Jackson75};
the algebra is tedious, but straightforward. The results for equations
\eq{Uncoupled} and \eq{BMT} are
\beqn{UnwrapForce}
\frac{d\vv}{dt}=\frac{q}{\g m}\vE'' +\frac{\geff}{\g m}\vB''
  + \frac{\Theta}{\g^2 m}\vs'
\eeqn
and
\beqn{SpinUnwrapGen}
\frac{d\vs}{dt}=\vs \cross \vect{\Omega}_{\text{new}},
\eeqn
where
\begin{eqnarray*}
\vect{\Omega}_{\text{new}}= \braces{\z - \frac{\g-1}{\g} \frac{q}{m}}\vB
  - \braces{\z -\frac{\g}{\g+1} \frac{q}{m}} \vv \cross \vE
  - \frac{\g}{\g+1} \braces{\z-\frac{q}{m}} \paren{\vv \vdot \vB} \vv \nline
  \dummy + \frac{\g-1}{\g}\frac{\geff}{m} \vE
  + \frac{\g}{\g+1}\frac{\geff}{m} \vv \cross \vB
  - \frac{\g}{\g+1}\frac{\geff}{m} \paren{\vv \vdot \vE} \vv
  + \frac{\Theta}{m(\g+1)} \vv \cross \vs
\end{eqnarray*}
and we have defined the convenient quantities
\begin{eqnarray*}
\z&\id&\frac{\m}{s}, \\
\Theta &\id&\z^2 \paren{\vE \vdot\vB} ,\\
\pard'&\id&\frac{\pard}{\pard t}+\frac{\g}{\g+1}\paren{\vv\vdot\del}, \\
\vs'&\id&\vs-\frac{\g}{\g+1}\paren{\vv\vdot\vs}\vv ,\\
\vE' &\id&\vE+\vv\cross\vB-\frac{\g}{\g+1}\paren{\vv\vdot\vE}\vv ,\\
\vE''&\id&\vE+\vv\cross\vB-\paren{\vv\vdot\vE}\vv ,\\
\vB''&\id&\vB-\vv\cross\vE-\paren{\vv\vdot\vB}\vv ,\\
\mbox{and } \spacer\geff &\id& \z \paren{ \vs\vdot\del}
  +\g\z\paren{\vs\vdot\vv}\pard'
  - \g\z^2\paren{\vs\vdot\vE'},
\end{eqnarray*}
and, in all expressions, the partial derivatives act only on the
external field quantities $\vE$ and $\vB$.

For ease of comparison with the equations of motion in current usage,
we present the Lorentz force law in the same form as \eq{UnwrapForce},
\beqn{UnwrapLorentz}
\frac{d\vv}{dt}=\frac{q}{\g m}\vE'',
\eeqn
and, likewise, the precession frequency vector for the
Thomas spin equation (\cite{Jackson75}, p.~559):
\[
\vOm_{\text{old}}= \braces{\z - \frac{\g-1}{\g} \frac{q}{m}}\vB
  - \braces{\z -\frac{\g}{\g+1} \frac{q}{m}} \vv \cross \vE
  - \frac{\g}{\g+1} \braces{\z-\frac{q}{m}} \paren{\vv \vdot \vB} \vv.
\]
It can be seen that, as advertised, the Lorentz and Thomas equations
are contained completely in the new equations. However, several
new features are present in both the new force equation,
\eq{UnwrapForce}, and the new precession
frequency vector, $\vOm_{\text{new}}$.
Most obviously, the magnetic dipole force is now included in
\eq{UnwrapForce}, albeit somewhat obscured by the multitude
of ``convenient quantities'' introduced for typographical sanity.
A recognition of this expression may again
be obtained by taking the non-relativistic limit (first order in $\vv$,
ignoring
Thomas precession); \eq{UnwrapForce} then returns us to
\[
\frac{d}{dt}(m\vv)=q\paren{\vE+\vv\cross\vB}
  +\paren{\vmu\vdot\del}\paren{\vB-\vv\cross\vE}-\dot{\vmu}\cross\vE,
\]
which is, as noted earlier, the now generally-accepted
\cite{Barone73,Aharonov88,Goldhaber89,Aharonov84} dipole force expression.
There are, of course, numerous new subtleties of \eq{UnwrapForce}
that arise from relativistic kinematics; we shall, however, defer a more
exhaustive investigation of them to another place.

Turning, now, to the spin precession equation \eq{SpinUnwrapGen},
what may come as a surprise to some is that there are differences
between $\vOm_{\text{old}}$ and $\vOm_{\text{new}}$.
The reason, however, is simply found:
the Thomas precession effects in the spin
precession equation depend on the \e{acceleration} of the particle;
clearly, if we have obtained a more accurate force law, then these
changes will necessarily feed through to the spin equation as well.
It is interesting to note that the approximate nature
of the Thomas expression $\vOm_{\text{old}}$ was noted explicitly
by Thomas himself \cit{Thomas27}, and by Bargmann, Michel and Telegdi
in their re-derivation \cit{Bargmann59},
but that, in the intervening decades, this approximate nature
has been lost on
many practitioners, who mistakenly believe $\vOm_{\text{old}}$
to be an \e{exact} expression.

At this point, it is worthwhile commenting on the absence of any
presence of \e{radiation reaction} forces in the equations that
have been derived above.
Their omission can, in fact, be traced back directly to the (implicit)
assumption that the electromagnetic potentials and fields in the
Lagrangian \eq{FinalLag} are the \e{externally generated ones only}.
In the case of the electric charge interaction, it was recognised
by Lorentz \cit{Lorentz06}, and later clearly
explained by Heitler \cit{Heitler54}, that this is an incorrect assumption.
For full consistency, one must include the potentials and fields \e{of the
particle itself} in the Lagrangian, even though they appear, at first
sight,  to be hopelessly
divergent quantities.
If one proceeds with extreme care, one can show that the
effects on the equations of motion of this ``self-interaction'' are, in
fact, twofold: firstly, the addition of the (finite) radiation reaction terms
of the Lorentz--Dirac equation; and, secondly, the \e{dynamical} effects
explaining the electromagnetic self-energy contributions to the rest
mass of the
system, for whatever arbitrary charge distribution is assumed.
The beauty of this procedure is that it reveals the all-encompassing nature
of the Lagrangian; no further input is required to obtain the equations of
motion.

The success of the above procedure
clearly indicates that a similar procedure should be undertaken
for the additional magnetic moment interaction Lagrangian present
in \eq{FinalLag}.
Barut and Unal \cit{Barut89} have considered the dipole radiation reaction
question from the point of view of a semi-classical \e{Zitterbewegung}
model, but to our knowledge an exact treatment of this problem
in terms of the classical spin vector $\vs$ has not been performed.
This problem is one that we are currently investigating.
However, it is vastly more complicated than the electric charge case,
by virtue
of the inclusion of \e{rotatonal} degrees of freedom for the particle.
In addition, one already knows in advance that the electric charge
and magnetic moment \e{interact} in their respective radiative terms,
as is evidenced by the Sokolov--Ternov effect
\cit{Sokolov63} (the polarisation of electrons due to their emitted
synchrotron radiation),
spectacularly confirmed in the polarisation experiments at LEP
in recent years \cit{LEP91,LEP92}.
Any prospective solution of the combined radiation reaction equations
for a charged, spinning particle must therefore, at the very least,
reproduce this important result.

\newsect{Summary}{Conclusions}
It has been shown, in this paper, that it \e{is} possible to construct
a fully consistent, comprehensive, relativistic classical Lagrangian framework
for analysing the motion of spinning particles possessing both electric
charge and magnetic dipole moments.
The results obtained here are not revolutionary.
They encompass the well-known Lorentz force and \TBMT\ equations.
They provide a rigorous foundation for the magnetic dipole force law
currently believed to be the most appropriate for such particles.
They further integrate, seamlessly, this dipole force with the Lorentz
force and
\TBMT\ equations, in a fully relativistic way.

It should be noted that the results of this paper \e{agree}
with the lowest-order terms obtained in the
analysis of Anandan, based on the Dirac equation
\cite{Anandan89}---and, of particular note, the ``Anandan force''
proportional to $\vE\cross\paren{\vmu\cross\vB}$,
which is, of course, simply the zeroth-order term in the force term
$-\vmudot\cross\vE$ above, when it realised that $\vmudot$ is,
via the Thomas-BMT equation, proportional to $\vmu\cross\vB$ to
zeroth order.
The current work, however, includes terms to \e{all} orders in the
particle's velocity, not just the lowest-order limit of Anandan.
It should also be noted that the criticisms of Casella and Werner
\cite{Casella92} of Anandan's analysis \cite{Anandan89} are
\e{erroneous}, being based on an obvious omission of all
``spin-flip'' terms from their quantum mechanical equations of motion.

The framework outlined in this paper
may now be used as a platform for full inclusion of
radiation reaction---not just for the electric charge, but also for
the magnetic moment, and their mutual interactions---in the classical limit.
If the enormous assistance provided by the existing classical radiation
reaction theory, both in terrestrial and astrophysical applications, is
any guide, then one can only speculate what additional richness of
physical phenomena will be made sensible with this addition to our
analytical resources.
\acknowledgments
Helpful discussions with I.\ Khriplovich, J.\ Anandan, N.\ Mukhopadhyay,
L.\ Wilets, P.\ Herczeg, A.\ G. Klein, G.\ I.\ Opat, J.\ W.\ G. Wignall,
A.\ J.\ Davies, G.\ N.\ Taylor,
K.\ Jones, R.\ E.\ Behrend and S.\ Bass are gratefully acknowledged.
This work was supported in part by the Australian Research Council,
an Australian Postgraduate Research Allowance
and a Dixson Research Scholarship.
We warmly thank the Institute for Nuclear
Theory at the University of Washington for its hospitality and the
United States Department of Energy Grant~\#DOE/ER40561 for partial support
during the completion of this work.

\end{document}